\documentclass[aip,jcp,reprint]{revtex4-1}
\usepackage[T1]{fontenc}
\usepackage[utf8]{inputenc}
\usepackage{graphicx}
\usepackage{amsmath}
\usepackage{amssymb}
\usepackage{color} 
\usepackage{pifont}
\usepackage{natbib}
\usepackage[left,columnwise]{lineno}
\usepackage{xcolor}

\begin{document}

\title{Competing Supramolecular Structures: Dielectric and Rheological Spectroscopy on Glycerol/Propanol Mixtures}

\author{Jan Philipp Gabriel}
\email{jan.gabriel@dlr.de}
\affiliation{Glass and Time, IMFUFA, Department of Science and Environment, Roskilde University, Roskilde, Denmark}
\affiliation{Institute of Material Physics in Space, German Aerospace Center, 51170 Cologne, Germany}


\begin{abstract}
Significant progress has been made in recent years in understanding the dynamics of pure hydrogen-bonded systems by analyzing the spectral shape of various susceptibilities\cite{Boehmer2014,bohmer2024dipolar,bohmer2025spectral}. Monohydroxy- and polyalcohols are currently considered to form transient supramolecular hydrogen-bonded structures in the form of chains, rings, and networks. This complex dynamic behavior has been identified in network-forming glycerol and chain-forming propanol by combining dielectric and light-scattering spectra. We apply these concepts to study the combined dielectric and shear rheological spectral shape of glycerol/propanol mixtures. Glycerol differs from propanol by having two additional hydroxy groups, which leads to significant differences in melting temperatures($\Delta T_{\textbf{m}}$\,=\,291\,K\,-\,147\,K\,=\,143\,K) and glass transition temperatures ($\Delta T_{\textbf{g}}$\,=\,190\,K-\,98\,K\,=\,92\,K). The strong difference results in two distinct calorimetric glass transitions at a molar glycerol concentration of $\chi_{gly}=0.3$, as well as a change in the shear modulus $G_{\infty}$ between $\chi_{gly}=0.5$ and 0.7. Performing a comprehensive analysis of the three applied experimental techniques leads to the conclusion that dielectric spectroscopy monitors the evolution of supramolecular chain and network structures and that the mechanical properties depend heavily on the formed hydrogen-bonded network. A strong dynamical heterogeneity is observed and manifests itself in two distinguishable glass transitions in dielectric spectroscopy and calorimetry. The presented chain/network mixture is dynamically highly heterogeneous when compared to the rather narrow dynamical heterogeneity in the network/network mixture Water/Glycerol.
\end{abstract}

\maketitle


\section{Introduction}
Hydrogen bonding liquids are complex, and their dynamic behavior is crucial for the function of biological systems and the formation of dynamic structures. Two of the intensively investigated substances in the literature apart from water are propanol \cite{kono1966comparison,Hansen1997, Schiener:1995, wieth2014dynamical, Gabriel:2017a, Sillren:2013} and glycerol \cite{davidson_glycerol_1951, Beevers:1980, Wuttke:1994, Yee:1996, Paluch:1996, Schneider:1998, chelli1999glycerol1, chelli1999glycerol2, Doess:2001, Vogel:2001, Ryabov:2003, Hensel:2004, Brodin:2005a, Gainaru:2008, Pronin:2010, Towey:2011, Meier:2012, Klieber:2013,sillren2014liquid, Gupta:2015, jakobsen2016thermalization, jensen2018slow, Kremer:2018, Niss:2018, Flaemig:2019, Gabriel2020a, gabriel2021molecular, bohmer2022glassy, henot2023orientational, arrese2025correlation}.

\begin{figure}[t!]
\centering
\includegraphics[width=0.99\linewidth]{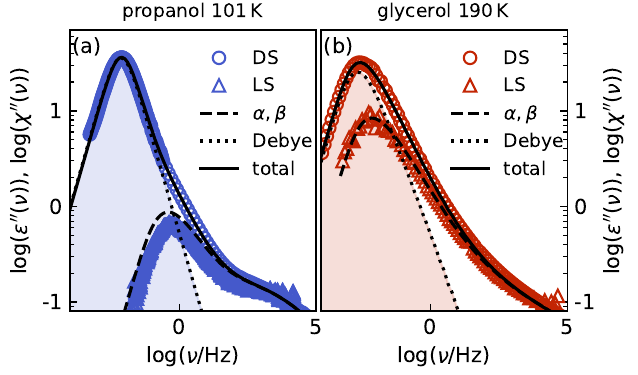}
\includegraphics[width=0.99\linewidth]{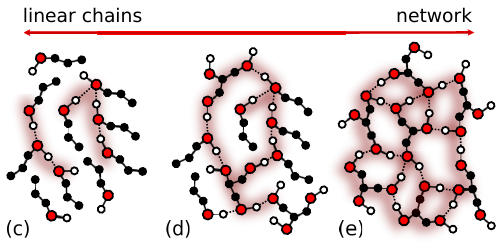}
\caption{
(a) Propanol \cite{Gabriel:2017a} and (b) glycerol \cite{Gabriel2020a} data from dielectric and light scattering show Debye, $\alpha$, and secondary relaxation processes. (c-e) Schematic representations illustrate the hydrogen-bonded chain and network structures in the propanol/glycerol mixtures.
}
\label{FIG1}
\end{figure}

As a prototypical monohydroxy alcohol (mono-alcohol), propanol has been the subject of scientific controversies for decades \cite{Boehmer2014}. Most mono-alcohols display the so-called Debye relaxation in their dielectric spectra, first observed by Debye\cite{Debye1929a} (cf. Fig.\,\ref{FIG1} (a)). This relaxation is slower than the $\alpha$ (structural) relaxation. The Debye relaxation stands out in BDS due to its enormous relaxation strength and its unusual mono-exponential shape. Currently, it is believed to originate from transient supramolecular hydrogen-bonded (H-bonded) structures \cite{gainaru2010nuclear}. Its large dielectric relaxation strength can be explained considering that the summation of permanent dipole moments of the bonded molecules results in a large effective dipole moment of the end-to-end vector of the chain\cite{gainaru2010nuclear} (see Fig.\,\ref{FIG1} (c)). The resulting cross-correlations are commonly quantified by the Kirkwood correlation factor $g_k$\cite{Kirkwood:1939a}, which is found to be larger than unity for most mono-alcohols. As the relaxation of the spatially extended end-to-end vector dipole moment averages over the local dynamic heterogeneities \cite{anderson1967molecular}, this leads in most cases to a mono-exponential relaxation, the Debye peak in the dielectric loss spectrum. The discussion is ongoing whether the hydroxyl group is involved in the Debye, $\alpha$, and possible intermediate processes  \cite{sillren2014liquid,arrese2020signature,becher2021molecular,cheng2024hydrogen,cates1987reptation}. 

Apart from chains, different supramolecular structures, such as 'brushes' \cite{Sillren2012}, and rings \cite{Dannhauser1968a, Singh2012}, can be formed in some mono-alcohols, resulting in different collective dipole moments. Which supramolecular structure is favored depends on the shape of the molecule, local steric restrictions, and temperature\cite{Bras2008, Dannhauser1968a}. The dielectric Debye contribution has long not been observed in mechanical and light scattering measurements of mono-alcohols\cite{Boehmer2014}, but finally observed in mechanical measurements as a bimodality resulting in a delayed terminal flow  \cite{gainaru2014shear,hecksher2014communication,hecksher2016communication,arrese2018multimodal,bierwirth2018communication,arrese2020signature} and in light scattering for some molecular liquids as an additional slow contribution\cite{Gabriel:2018a}.

For a long time, the dielectric signal of polyhydric alcohols (poly-alcohols) was considered to consist only of one structural relaxation peak \cite{bohmer2022glassy}. For glycerol, a similar dielectric cross-correlation contribution as in mono-alcohols can be seen in the dielectric spectrum in Fig.\,\ref{FIG1} (b)\cite{Gabriel2020a}, which we assume results from hydrogen-bonded network structures resulting in schematically illustrated in Fig.\,\ref{FIG1} (e) and recently confirmed in simulations\cite{henot2023orientational}. The mechanical spectrum of glycerol demonstrates a bimodal peak in the shear modulus\cite{jensen2018slow,arrese2020signature} and shear compliance\cite{Gabriel2020a} connected to the Debye and $\alpha$ relaxation identified in dielectric and light scattering experiments \cite{Gabriel2020a}.

An alternative to disentangle dynamics by cooling is mixing strongly dynamic asymmetric substances. A polymer-softener system like methyl-tetrahydrofuran (MTHF) and oligomeric methyl methacrylate (OMMA) has a $T_{\textbf{g}}$ separation of $\Delta$T$_\textbf{g}=249$\,K. In this system, two glass transitions are observed by disentangling the dynamics with photon correlation spectroscopy and x-ray photon correlation spectroscopy \cite{schramm2010concentration}. A similar experiment was performed with polymethyl methacrylate (PMMA) and Picolin \cite{boehmer2021origin} with a $T_{\textbf{g}}$ separation of $\Delta$T$_g=227$\,K. These results resolved separate dynamics and glass transition temperatures by light scattering, dielectric spectroscopy, MD simulations, \cite{boehmer2021origin}, and NMR \cite{korber2021reorientational}.

The mixture of propanol and glycerol was investigated in dielectric (50\,Hz-600\,kHz) and mechanical spectroscopy (10-130\,MHz) by Kono et al. in 1966 \cite{kono1966comparison}. At that time, the dielectric spectroscopy of alcohols was not assumed to represent structural relaxation, and the idea of connecting the strong relaxation of alcohols with transient hydrogen-bonded superstructures was not common \cite{gainaru2010nuclear}. Later, the structural relaxation of propanol was isolated by Hansen et al. with combined dielectric and light scattering experiments \cite{Hansen1997}. Recently, the full spectral shape and spectral identification of hydrogen-bonded cross-correlated structures were analyzed for propanol \cite{Gabriel:2017a} and glycerol \cite{Gabriel2020a} by combining dielectric spectroscopy and light-scattering. In both substances, it was shown that slow cross-correlations dominate the dielectric loss spectrum as illustrated in Fig.\,\ref{FIG1} (a) and (b). The self-correlation representing $\alpha$ and secondary relaxation of the molecules is measured by depolarized dynamic light scattering and thus, can be identified in the dielectric spectrum.

A large number of mono-alcohols with strong and weak dielectric signals or Debye processes were mixed and summarized in a very comprehensive manner by Bierwirth et al. \cite{bierwirth2018mixturealcohol}, discussing concepts of classical binary mixing, ideal mixing, closing and opening of chains and rings. Most of them mix well and show in their dielectric spectra different kinds of changes in their spectral shapes and intensities. The most agreed-upon explanation is a change in hydrogen bonding structures caused by mixing, leading to dilution and changing cross-correlation. An example from this huge treasure trove is the mixture of the mono-alcohol 2-ethyl-1-hexanol (2E1H) and the non-alcohol 2-ethyl-1-hexylbromide (2E1Br). Despite only having a small $\Delta$T$_g=13$\,K, they show a huge heterogeneous dynamic caused by the separated dynamics of hydrogen-bonded structures and not hydrogen-bonded structures. In mono alcohols, usually the hydrogen-bonded network is not calorimetrically active as shown for 2E1H \cite{huth2007comparing}. The 2E1H/2E1Br mixture is less heterogeneous than the investigated propanol/glycerol mixture, and the hydrogen-bonded network is diluted by the alcohol component and eliminating the origin of the heterogeneity of the system.

The structure and dynamics of hydrogen-bonded associations found in propanol and glycerol are markedly different. The focus will be on the evolution of the dynamics upon mixing. Therefore, combining mechanical and dielectric spectroscopy on propanol/glycerol mixtures allows for analyzing the macroscopic and molecular behavior, which is shown to be complex. We pursue the question of whether the dielectric chain and network signature identified in the pure components is still present in the mixture. Furthermore, we aim to study the impact of mixing on the mechanical properties of the hydrogen-bonded liquids. We begin by analyzing calorimetric data of the mixtures and subsequently present dielectric spectra, followed by mechanical spectra. The discussion will include a comparison of the three methods in the context of the literature on molecular dynamics in binary mixtures, and give remarks on demixing effects seen by X-ray.


\section{Method}
All samples were mixed and loaded in a glovebox and then transferred into the cryostat with minimal air contact. The propanol was bought at Alfa Aesar with a purity of 99.9\%, and the glycerol was bought from Alfa Aesar with a purity of 99.5\%. Mixtures were shaken by hand and the intermediate concentration was kept overnight at 70$^\circ$C to obtain a homogeneously mixed sample. All samples were fully mixed and transparent at room temperature and showed no visible demixing over months. 

The dynamic shear modulus was measured over more than six decades of frequency (10\,mHz to 80\,kHz) with a three-disc piezoelectric shear gauge \cite{christensen_rheometer_1995, hecksher2013mechanical, mikkelsen_piezoelectric_2022}. This technique is optimal for measurements on stiff systems (1\,MPa to 10\,GPa)\cite{christensen_rheometer_1995} and is well-suited for study liquids around their glass transition temperature.

Dielectric measurements were performed in a parallel plate capacitor placed inside the same custom-built cryostat system as used for the mechanical measurements\cite{igarashi_cryostat_2008, igarashi_impedance-measurement_2008}. By this, we ensure that shear and dielectric data sets can be measured under identical experimental conditions.

\begin{figure}[t!]
\centering
\includegraphics[width=0.99\linewidth]{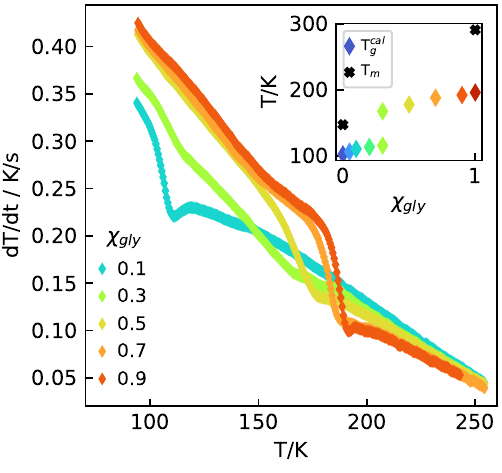}
\caption{
Selection of calorimetric measurements of glycerol/propanol mixtures showing two calorimetric features at 50\% mol. The inset shows the obtained onset glass transition temperatures $T_{\textbf{g}}$ and melting temperatures of the pure substances.
}
\label{FIG2}
\end{figure}

For the calorimetry, small glass test tubes of 0.5 ml were prepared in a glove box and quenched in liquid nitrogen. The change of temperature over time was monitored while the sample heated up in a Styrofoam-insulated environment. In this method, the cooling and heating rates change over the whole temperature window. It is assumed that $dT/dt$ is inversely proportional to the heat capacity $C$ of the sample. With $R_0$ as the thermal resistance of the box and $T_0$ the target temperature, we assume:
\begin{equation}
\frac{dT}{dt} = \frac{1}{R_0 C} (T(t)-T_0).
\end{equation}
The calorimetric method used is described in more detail in \cite{jakobsen2016thermalization}.


\section{Results}

As a first characterization, we consider the calorimetric response of the propanol/glycerol mixtures in Fig\,\ref{FIG2}, showing the time-dependent temperature change while the sample is continuously heated after being quenched to 77\,K in liquid nitrogen. In propanol and glycerol-rich concentrations, one glass transition step is identified in the heating curves. The glass transition temperatures $T_g$ are obtained by determining the onset temperature of the glass step. As shown in the inset, $T_g$ changes with varying concentrations, following two continuous parallel lines, one reflecting propanol-rich and the other glycerol-rich concentrations. At the intermediate concentration $\chi_\mathrm{gly}=0.3$, two distinct kinks, interpreted as glass transition steps are observed. The concentration-dependent $T_\mathrm{g}$ changes linearly in parallel starting from the pure propanol and glycerol $T_\mathrm{g}$. 

The observation of two glass transition temperatures is typical for high $T_{\textbf{g}}$ contrast mixtures \cite{schramm2010concentration, boehmer2021origin} or could be connected to an unknown type of phase separation. Visually monitoring the samples reveals no signs of demixing on the length scales of visible light while heated from liquid nitrogen, making domains larger than some hundred nanometers (the light wavelength) unlikely. Despite this, not yet published X-ray experiments (with Dorthe Posselt at RUCSAX) show structural changes larger than the molecular structures at low temperatures. 

\begin{figure*}[t!]
\centering
\includegraphics[width=0.99\linewidth]{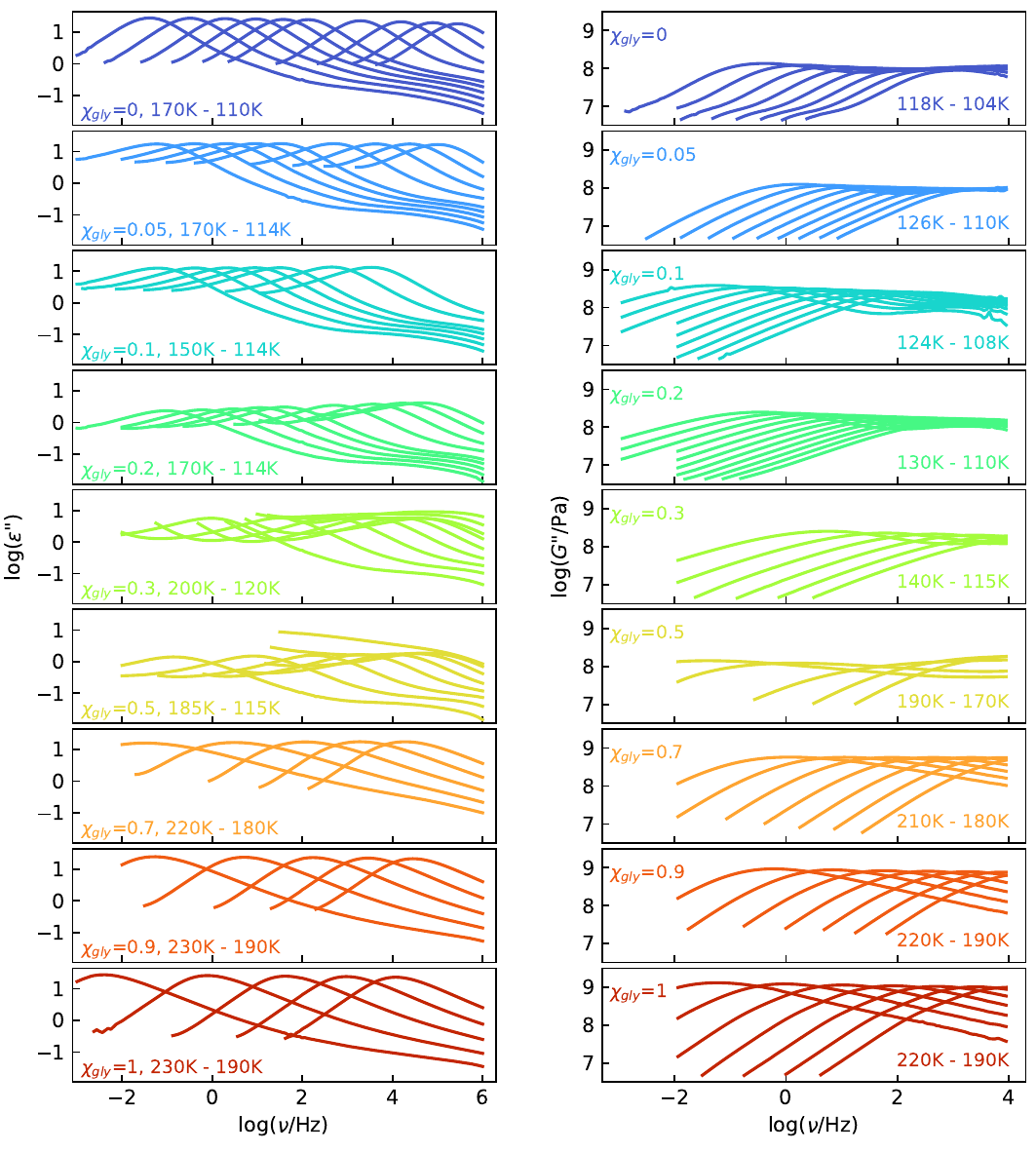}
\caption{
Concentration-dependent panels show dielectric loss spectra (left) and mechanical shear modulus (right) between 110K and 230K. Dielectric spectra are cut at the frequency where the DC conductivity starts.
}
\label{FIG3}
\end{figure*}

The measured dielectric and mechanical loss spectra for different concentrations and temperatures are presented in Figs.\,\ref{FIG3}. Full loss and storage spectra at all temperatures are presented in the supplementary material. All spectra have in common that with decreasing temperature, the position of the main peak shifts from high to low frequencies, and the spectral shapes of the pure components broaden upon mixing. At frequencies lower than the main peak, the dielectric spectra show conductivity covering any potential lower-frequency relaxation processes. For better visualization, we limit the presentation of the spectra in Fig.\,\ref{FIG3} to frequencies where the DC conductivity is weak. 


\section{Analyses}

\begin{figure}[t!]
\centering
\includegraphics[width=0.99\linewidth]{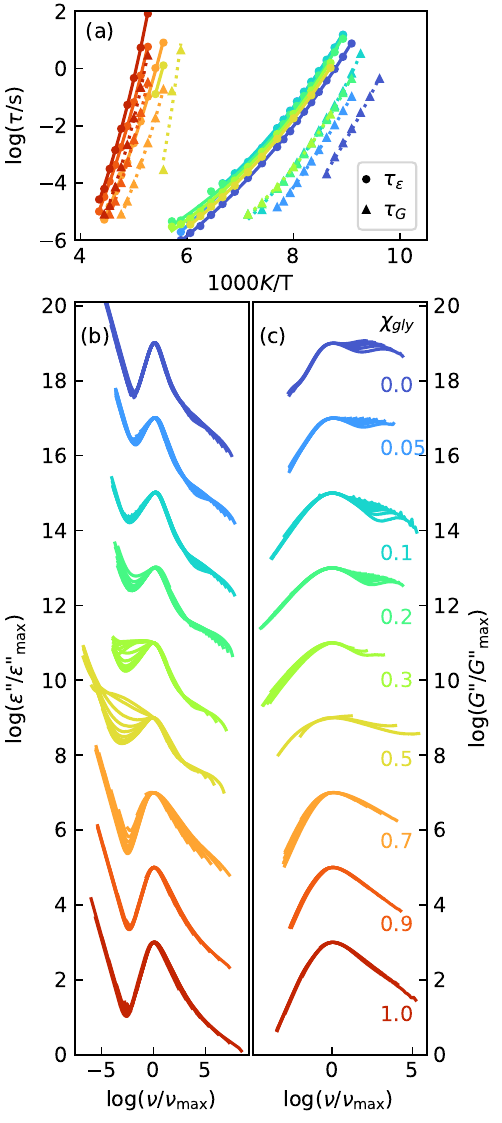}
\caption{
(a) Concentration-dependent main peak time constants extracted from shear and dielectric spectra by peak picking at identifiable peak positions. Dashed lines are VFT fits. Normalized master plots to main peak frequencies of concentration-dependent (b) shear and (c) dielectric spectra.
}

\label{FIG4}
\end{figure}

For the dielectric spectra of the mixtures, only one main relaxation peak can be identified, except at a few temperatures and concentrations, where very broad or even bimodal peaks are observed. In Fig.\,\ref{FIG4} (b,c) we plot the same spectra as in Fig.\,\ref{FIG3} with the frequency axis being normalized to the peak frequency $\nu_{max}$ to obtain one master plot for every concentration. The resulting master plot demonstrates that in glycerol-rich mixtures, the main peaks, especially in the dielectric spectra, obey time-temperature superposition (TTS). For propanol-rich concentrations, only the main peak and the low-frequency behavior show TTS, and the high-frequency spectra are similar in shape to the pure propanol spectra. For most concentrations, the conductivity contribution to the dielectric spectrum collapses well with this scaling and masks potential slower relaxation dynamics. Additionally, for increasing glycerol concentration, a broadening of the low-frequency flank of the main peak is observed. Above a concentration of $\chi_{gly} = 0.5$ no main peak (originating from propanol-like dynamics) can be distinguished, and the spectrum is narrowing again (in the range of glycerol-like dynamics). All deviations from TTS remaining in the dielectric spectrum should arise from different temperature dependencies of the underlying processes (secondary relaxation vs. $\alpha$ relaxation). The mechanical spectra display similar TTS behavior to the dielectric spectra. Like in the dielectric spectra, the mechanical spectra are extremely broadened around $\chi_{gly} = 0.5$ and appear to separate into two contributions despite not showing well-distinguishable peaks.

The temperature-dependent peak time constants $\tau_{max} = 1/(2\pi\nu_{max})$ used for creating the master plots for dielectric and mechanical spectra are shown in Fig.\,\ref{FIG4} (b,c) and are in all cases well-described by the Vogel Fulcher Tammann (VFT) equation.

\begin{equation}
\tau_{\text{VFT}} =
\tau_0 \exp \left( \frac{\Delta E}{T(t)-T_{\text{VFT}}} \right)
\label{eqvft}
\end{equation}

The observed temperature dependencies in Fig.\,\ref{FIG4} (a) can be divided into propanol-like and glycerol-like. With increasing glycerol concentration, the dielectric time constants (circles) shift initially a bit but then stay nearly the same. The dielectric time constants of glycerol-rich concentrations slow with increasing propanol concentration. Only at $\chi_{gly} = 0.5$ can two peak time constants be extracted from the dielectric spectra (of approximately $1$\,s and $10^5$\,s). The temperature dependency of the faster peak time constants is consistent with time constants of glycerol at slightly higher glycerol concentrations, while the faster time constant is consistent with the propanol dynamics at higher propanol concentrations.

\begin{figure}[t!]
\centering
\includegraphics[width=0.99\linewidth]{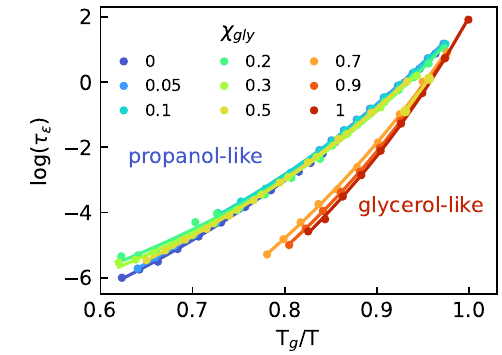}
\caption{
Temperature-dependent dielectric correlation times scaled to the dielectric glass transition temperature.
}
\label{FIG10}
\end{figure}

The dielectric time constant can be further analyzed by scaling it to the respective dielectric glass transition temperature at 100\,s as shown in Fig. \ref{FIG10}. The surprisingly good collapse in a propanol-like and glycerol-like temperature dependency. The change occurs between $\chi_{gly} = 0.7$ and 0.5. It is remarkable that at the concentration $\chi_{gly} = 0.5$ both observed peaks follow the continuation of the propanol-like and glycerol-like temperature dependence.

The time constants extracted from the mechanical modulus (triangles) appear to have similar temperature-dependency as their dielectric counterparts, but are 0.5 to 4 orders of magnitude faster, depending on concentration. These are remarkably huge differences. While at $\chi_{gly} = 0.5$, two processes are visible in dielectric spectroscopy; the mechanical spectra are dominated by the glycerol-like contribution, while the propanol-like seems to appear less prominent than in the dielectric spectrum. At higher glycerol concentrations, both methods show similar time constants quickly approaching values like pure glycerol. In the mechanical measurements, only pure glycerol is approaching a terminal flow as shown in more detail in the supplementary material, Fig. 1 and 2, inspired by \cite{bierwirth2018communication}.

\begin{figure}[t!]
\centering
\includegraphics[width=0.99\linewidth]{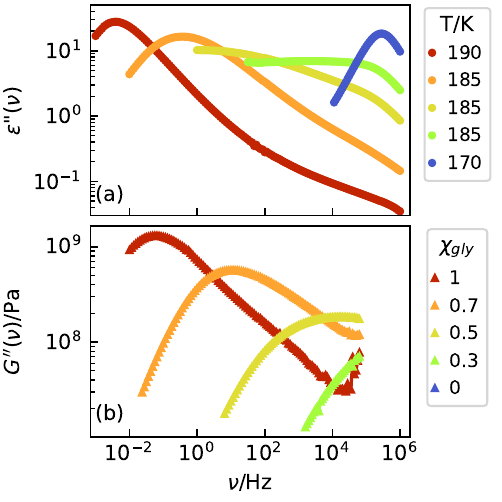}
\caption{
(a) Dielectric permittivity and (b) mechanical shear modulus of propranolol/glycerol mixtures with the variation of glycerol concentration $\chi_{gly}$ in a narrow temperature range between 170K and 190K.
}
\label{FIG5}
\end{figure}

Since the mixtures are strongly dynamically asymmetric ($\Delta T_{\text{m}} = 291\,K-147\,K = 143\,K$, $\Delta T_{\text{g}} = 190\,K-98\,K = 92\,K)$) it is hard to cover the whole dynamic and temperature range. This is difficult due to a limited spectral range, and further limited in dielectric spectroscopy by conductivity. For showing the whole relaxation behavior in the full concentration range, a narrow temperature window of 170\,K to 190\,K is selected and the dielectric permittivity is presented in Fig.\,\ref{FIG5} (a). The heterogeneous dynamics of both components cover 8 orders of magnitude in frequency within a 20\,K temperature interval. At intermediate concentration $\chi_{gly} = 0.3 - 0.7$ and fixed temperature 185\,K, the dynamics changes from a broad spectrum spanning from glycerol-like to propanol-like peaks. At $\chi_{gly} = 0.5$ two separated contributions can be identified.

The mechanical spectrum presented in Fig.\,\ref{FIG5} (b) has a narrower spectral window than the dielectric spectroscopy, ranging from 10\,mHz to 80\,kHz, and shows less clearly separated processes. The propanol-rich concentrations are too slow to have a visible component. In glycerol-rich concentrations $\chi_{gly}=0.7$ to $1$, we can not see the bimodality of pure glycerol at low frequencies from higher temperatures, but the spectra become broader with increased glycerol concentration, like in the dielectric up to $\chi_{gly}=0.5$. With increasing glycerol content, the spectrum becomes narrower, and the dynamics change continuously. 

\begin{figure}[t!]
\centering
\includegraphics[width=0.99\linewidth]{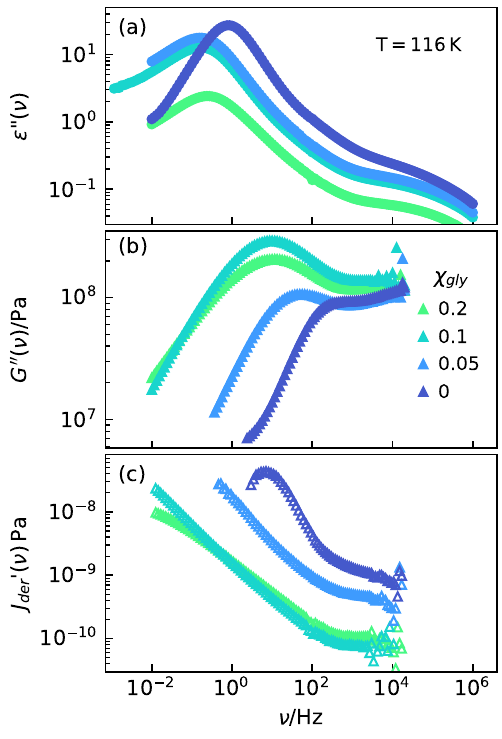}
\caption{
(a) Dielectric permittivity loss spectra and (b) mechanical shear modulus loss spectra at 116\,K. (c) Derivatives of the shear mechanical compliance storage part, definition see eq.\,\ref{eqJder}.
}
\label{FIG6}
\end{figure}

Comparing imaginary parts of dielectric and mechanical spectra at a fixed temperature and low glycerol concentrations between $\chi_{gly}=0$ and 0.2 shows that the characteristic dielectric spectrum of propanol is visible in Fig.\,\ref{FIG6} (a). The main peaks broaden with increasing glycerol concentration, but cannot be monitored at lower frequencies, and the data are cut when the conductivity appears. The dielectric main peak frequency hardly changes with increasing glycerol concentrations. The mechanical spectrum (see Fig.\,\ref{FIG6} (b)) displays a significantly stronger influence of the increase in glycerol concentration. Generally, the identified mechanical main peak is faster than the dielectric one. Since the dielectric spectrum is strongly influenced by the Debye process, or in other words, by hydrogen structures. In the mechanical modulus spectrum, Debye, $\alpha$, and $\beta$ cannot be easily disentangled. In both methods, a comparable $\beta$-process is visible, but no Debye and $\alpha$ process can be distinguished without additional information by light scattering, for example.

\subsubsection{Shear compliance}
In pure glycerol, we showed that in the compliance representation, shear rheology and dielectric spectroscopy have the same spectral contribution at the Debye and $\alpha$ time scale \cite{Gabriel2020a}. Therefore, it is interesting to obtain compliance with the main peak time constants and to know if there is also a separable propanol $\alpha$ process time constant in the mechanical spectrum. In glycerol, the comparison was done by using the concept of generalized susceptibilities $\chi''(\omega)$ and comparing light scattering, dielectric spectroscopy, and NMR susceptibility with the shear compliance representing a mechanical susceptibility\cite{Gabriel:2017a,Gabriel2020a}.

Since we cannot identify a pure flow contribution like in the pure components, we apply a method well-established in dielectric spectroscopy. Instead of subtracting the flow part for obtaining the recoverable compliance \cite{Verney1989} as generalized compliance $J''(\omega)-J_{\text{flow}}=\chi''_{\text{J}}(\omega)$  we use the derivative method \cite{wubbenhorst2002analysis} by extracting an approximation (broad spectra limit) for the recoverable compliance.

\begin{equation}
J'_{\text{der}}(\omega) = -\frac{\pi}{2}\frac{dJ'(\omega)}{dln(\omega)} \approx \chi''_{\text{J}}(\omega)
\label{eqJder}
\end{equation}

The approximation for Debye shaped processes \cite{richert2011appearance,roy2014dielectric} is not applicable for the broad shear spectra. For propanol rich mixtures the recoverable compliance estimate $J'_{\text{der}}(\omega)$ is shown in Fig.\,\ref{FIG6} (c). Only propanol is reaching a maximum, and no clear low-frequency behavior can be extracted. The main peak in the compliance representation is slower than the dielectric $\alpha$ relaxation peak but faster than the Debye contribution.

\begin{figure}[t!]
\centering
\includegraphics[width=0.99\linewidth]{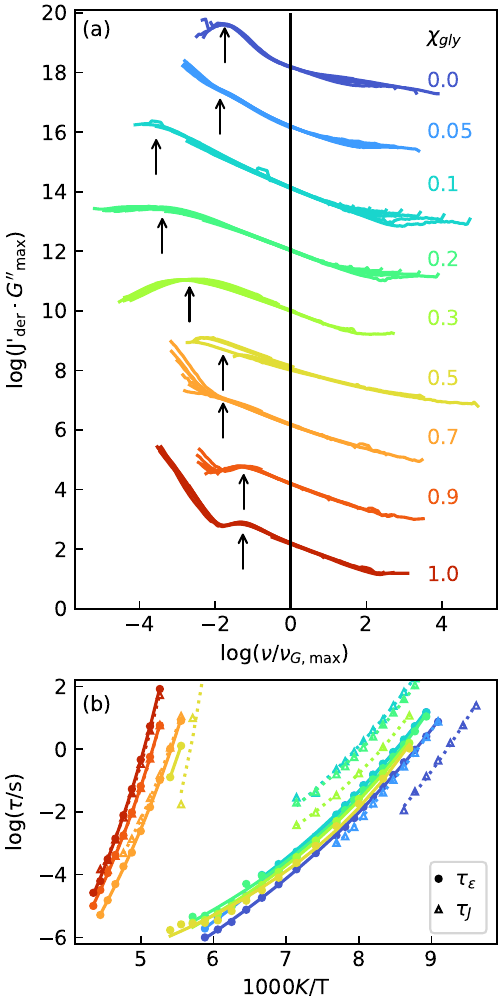}
\caption{
(a) Concentration-dependent master curves of shear compliance spectra normalized to the main shear modulus peak. (b) Concentration-dependent main peak time constants extracted from shear compliance and BDS spectra at identifiable peak positions. Dashed lines are VFT fits eq.\,\ref{eqvft}.
}
\label{FIG7}
\end{figure}

Extracting compliance time constants is difficult because the sensitivity in the decisive region is decreasing dramatically. We utilize the master plot normalization $G_{\text{max}}$ and $\nu_{\text{max}}$ from the modulus representation and convert the shear mechanical data to the normalized $J'_{\text{der}}(\omega)$ as an estimate for the recoverable compliance presented in Fig.\,\ref{FIG7} (a). Black arrows mark an approximation for the mechanical compliance main peak positions. The point where the curves are departing from the master plot behavior at low frequencies indicates the resolution limit. For the pure and some intermediate concentrations, this works well. 

The dynamic separation between the indicated main peak and the normalization frequency $\nu_{\text{G,max}}$ in Fig.\,\ref{FIG7} (a) allows us to estimate $\tau_{\text{J,max}}$, which we present in Fig.\,\ref{FIG7} (b). The mechanical compliance time constants obtained in this way are slower than the modulus time constant and, at high glycerol concentration, strongly correlated to the time constant of the dielectric Debye process. In intermediate glycerol concentrations the compliance time constant cannot be obtained precisely, but it is clear that the largest dynamic separation between $\tau_{\text{G,max}}$ and $\tau_{\text{J,max}}$ is around $\chi_{gly} = 0.2$, where it is found to be more than four magnitudes slower than the modulus time constant. In this regime, a comparison with dielectric is difficult because conductivity covers the relaxation times of dielectrics related to the observed process. The compliance peak of $\chi_{gly} = 0.3$ is the most pronounced and broadest, but also shows that it is not the slowest mode in the mechanics, despite being much faster than the slowest dielectric mode. At propanol-rich concentrations, the compliance time constant evolves from slower than the dielectric Debye peak to faster but still slower than the propanol $\alpha$ relaxation. Mechanical time constants are plotted relative to the mechanical in the supplementary material Fig.\,3.

\subsubsection{Mechanical Shear Modulus $G_{\infty}$}

\begin{figure}[t!]
\centering
\includegraphics[width=0.99\linewidth]{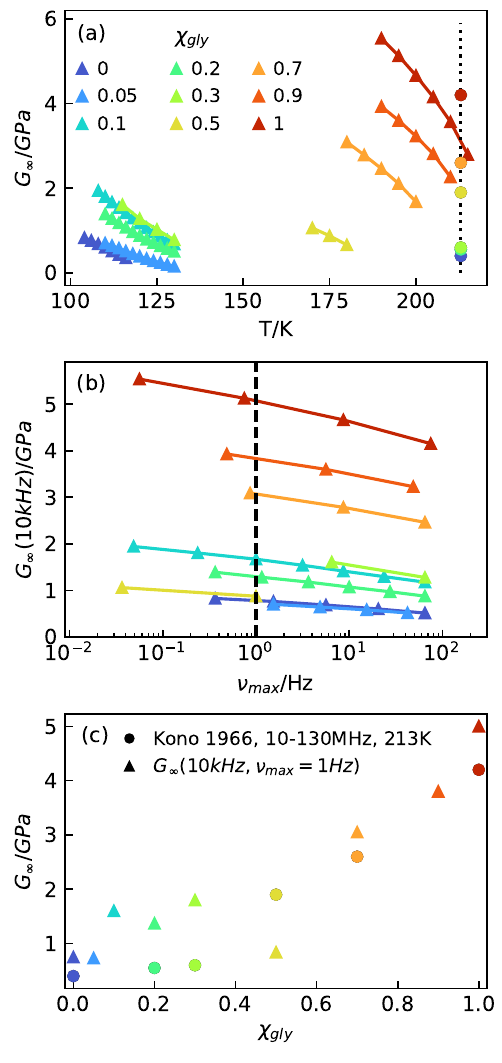}
\caption{
$G_{\infty}$ at 10\,kHz over (a) Temperature and (b) Relaxation Peak frequency of 1Hz close to $T_G$. In (c) shows $G_{\infty}(10kHz)$ close to $T_g$ collected at $\nu_{max}$=1Hz (dashed line (b))  and $G_{\infty}$ at 10MHz and 213K (doted lines (b)) by Kono et al. \cite{kono1966comparison}.
}
\label{FIG8}
\end{figure}

The mechanical stiffness of the mixture is analyzed in Fig.\,\ref{FIG8} (a) with the temperature-dependent high-frequency plateau of the real part of the mechanical shear modulus $G_{\infty}$. To make this comparable, $G_{\infty}$ at 10\,kHz is plotted for all concentrations over the main relaxation peak frequency of the mechanical spectra, as shown in Fig.\,\ref{FIG8} (b). High concentrations of mixtures continually soften with increasing propanol concentration. However, upon further dilution at concentrations between $\chi_{gly}$ 0.7 and 0.5, the mixture softens to the stiffness of pure propanol. In the Kono et al. \cite{kono1966comparison} data collected at higher frequencies as 212\,K between 10 and 130\,MHz, the softening might be found later between $\chi_{gly}$ 0.5 and 0.3. This effect might be a sign that increasing propanol content continuously weakens the hydrogen-bonded network of glycerol, by replacing interconnected glycerol structures with propanol chains until the network is no longer fully percolated at concentrations below 0.6
. The measurements themselves are quite precise in relative terms, but have an unknown absolute error around 1GPa. Since the results by Kono et al. are at higher frequencies, they are, in general, better suited to measure $G_{\infty}$ and might be less sensitive to filling uncertainties. Nevertheless, Kono's data are recorded at high temperatures of 213K for the dynamics. They are therefore difficult to compare to the analysis near $T_g$ (at 1Hz relaxation peak frequency) as in this work depicted in Fig.\,\ref{FIG8} (c).

\subsubsection{Transition Temperatures $T_{g}$ and  $T_{m}$}

\begin{figure}[t!]
\centering
\includegraphics[width=0.99\linewidth]{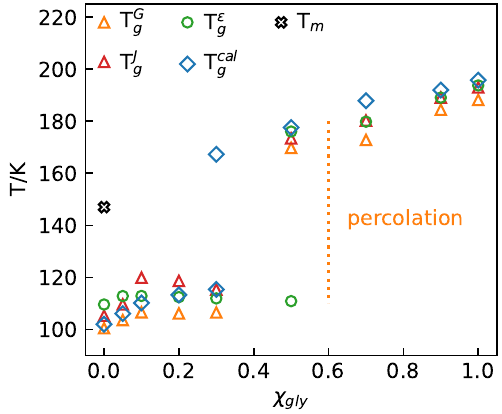}
\caption{
Glycerol-dependent calorimetric, BDS (evaluated at $\tau = 100$\,s), and shear-modulus (evaluated at $\tau = 100$\,s) glass transition temperatures and melting temperature of propanol and glycerol.
}
\label{FIG9}
\end{figure}

The dynamic of propanol and glycerol appears nearly independent when comparing T$_{\textbf{g}}$\,s in Fig.\,\ref{FIG2}. The dielectric and mechanical glass transition temperatures can be obtained by defining T$_g^{\epsilon}$, T$_g^{G}$ and, T$_g^{J}$ as the temperature where the VFT (see eq.\,\ref{eqvft}) fits from Fig.\,\ref{FIG4} (a) reach 100\,s ($T_{\textbf{g}}$) time constant in caloremetry quenched at high cooling is closer to 10s \cite{Boehmer:2019a}). Including the dielectric and mechanical glass transition temperatures in Fig.\,\ref{FIG9} gives a consistent picture with calorimetry T$_g^{cal}$. All methods show separate glycerol-like and propanol-like behavior. The glycerol-rich $T_{\textbf{g}}$s are all following the same behavior upon threshold between $\chi_{gly}=0.7-0.5$. This might be caused by a percolation phenomenon. Initially, at low glycerol concentrations, one observes propanol-like behavior in all methods; however, between $\chi_{gly}=0.2-0.5$, the methods exhibit different sensitivities in that concentration regime. The T$_g^{cal}$ (blue diamonds) concentration dependency follows separate changing propanol-like and glycerol-like behaviors. While the dielectric T$_g^{\epsilon}$ (green circles) follows both the calorimetric T$_g^{cal}$s (when observable). The shear modulus T$_g^{G}$ (orange triangular) and shear compliance T$_g^{J}$ (red triangular) are separating in propanol- and glycerol-like behavior in the range  $\chi_{gly}=0.1-0.3$. This might be a consequence of the different weighting of slow Debye and fast $\alpha$ relaxation contributions. The modulus T$_g^{G}$ is staying with the fast propanol dynamics, and the compliance T$_g^{J}$ is emphasizing the slow glycerol dynamics.

\section{Discussion}

The dielectric spectrum of pure propanol consists of three processes called from low to high frequency: Debye, $\alpha$, and $\beta$ processes \cite{Hansen1997, Gabriel:2017a}. Traditionally, the dielectric spectrum of glycerol is interpreted to contain the $\alpha$ process with an additional high-frequency excess wing. However, recent investigations showed that it can be interpreted similarly to propanol, consisting of a Debye and $\alpha$ process \cite{Gabriel2020a}. Long-aged \cite{schneider_glycerolbeta_2000} and hyper quenched pressure-densified glycerol \cite{gainaru2020suppression} have shown that below the excess wing is a separable $\beta$ relaxation similar to that of propanol.

As discussed in the introduction, the Debye process is assumed to originate from cross-correlations of dipole moments introduced by supramolecular structures. Therefore, the analyses were done from the perspective that in the mixtures the main peak contributions in the dielectric spectra will also arise from dipolar cross-correlation as illustrated for pure glycerol and propanol in Fig\,\ref{FIG1} (d) and (e), where light scattering isolates cross-correlation from self contributions in the dielectric signal. The cross-correlation assumption is supported by the observation of a stronger relaxation strength than expected for pure self-correlation at all concentrations (see fig. 4-6 in the supplementary material), resulting in Kirkwood factors between those of glycerol and propanol (Kirkwood factor depends on temperature and method of determination, at T$_g$ it is 2.5 for glycerol and 4.5 for propanol \cite{bohmer2024dipolar}). It is known that diluting systems with unpolar solvents destroys cross-correlation \cite{pabst2020dipole} and that structures with anti-correlations can form \cite{singh2012watching,Boehmer:2019a,bohmer2023revealing}, which are not observable with such a high Kirkwood-factor/dielectric-relaxation strength.

Self-correlations connected to molecules or parts of molecules measure changes in their dynamics under the influence of neighbouring molecules. In monohydric alcohols, the response to temperature changes is not on the time scale of the Debye relaxation of the superstructure; rather, it is on the time scale of the molecular self-correlations\cite{gabriel2025linear}. However, the relaxation of chains and rings can be affected by strong electric fields \cite{gabriel2023comparing}.

The propanol/glycerol mixture, as illustrated in Fig.\,\ref{FIG1} (c-e), changes with varying concentrations from being dominated by (a) propanol-like hydrogen-bonded chains to (b) a glycerol-like hydrogen-bonded network. In between (c), a transition occurs around the assumed percolation transition of the hydrogen-bonded network, both behaviors coexist. The mechanical stiffness quantified by $G_{\infty}$ (cf. Fig.\,\ref{FIG8}) indicates that the percolation transition occurs between $\chi_{gly} = 0.5$ and $0.7$. Kono et al. \cite{kono1966comparison} measured in 1966 $G_{\infty}$ as well and would find a percolation transition more between $\chi_{gly} = 0.3$ and $0.5$. If we assume that the glycerol network can be described by a bonded percolation on a cubic lattice already at a concentration of around 0.248, the system would be percolated \cite{stauffer2003introduction}. This oversimplification would lead to glycerol already forming a percolating network at $\chi_{gly} = 0.25$, which would be in volume percentage $\chi_{gly}^{vol} = 0.43$. 

The large dielectric relaxation strength at all concentrations indicates that cross-correlations dominate these mixtures, and the dynamics in dielectric spectroscopy are dominated by the dynamics of hydrogen-bonded structures. Below the percolation transition, between $\chi_{gly} = 0-0.5$, the predominance of propanol-like chain structure is manifested by the Debye peak in the dielectric spectrum (cf. Fig \ref{FIG6}(a)) and the remaining $\alpha$ and $\beta$ relaxation are nearly identical to pure propanol. With increasing glycerol concentration, the amplitude of the Debye contribution decreases, and a broadening of the main peak is observed. In the percolated regime around $\chi_{gly} = 0.7-1$, the dielectric spectrum displays glycerol-like behavior which is dominated by a slightly broadened cross-correlation peak (cf. Fig.\,\ref{FIG5} (a)) that is dynamically coupled to the shear compliance contribution as seen by identical time constants $\tau_{\epsilon}$ and $\tau_J$ in Fig.\,\ref{FIG7} (b). These two time constants display a separation below the percolation transition and different temperature dependencies. As the glycerol concentration decreases, the shear compliance time constants are decoupled from the dielectric time constants and continuously get faster. Finally, for pure propanol, the compliance time constant $\tau_J$ is faster than the Debye time constant $\tau_{\epsilon}$ representing hydrogen-bonded superstructure and only slightly slower than the $\alpha$ relaxation. In a similar system 1-butanol, NMR has found on this intermediate timescale between Debye and $\alpha$ relaxation\cite{sillren2014liquid}, the opening and closing of hydrogen groups \cite{gainaru2010nuclear}. This could mean that the dielectric $\alpha$ relaxation is coupled to the reorientation of propanol, the shear compliance to the opening and closing of hydrogen bonds, and the dielectric Debye process to the reorganization of supramolecular chain structures. 

As a naive picture, one could believe that on the one hand, propanol can have separated dynamics between simple reorientation and hydrogen breaking since it accepts only a maximum of two hydrogen bonds and donates to one, while glycerol, on the other hand, can accept up to six hydrogen bonds and donates to three hydrogen bonds. Having this in mind, the coupling between alpha relaxation, hydrogen breaking, and superstructure is much stronger in a network structure than in a chain structure. Network and chain-forming systems generate effective additional dipole cross-correlations. In this regard, chains are easy to handle by adding projections onto an end-to-end vector, but it can be more difficult to analyse the local preferred directions of networks. Glycerol tends to form hydrogen bonds in one direction and van der Waals bonds in the other. This results in lower cross-correlation intensity in glycerol than in propanol.

As indicated in Fig.\,\ref{FIG1} (a), the propanol chain structures result in a large separation between dielectric cross-correlation and the hydrogen chains from the $\alpha$ relaxation, while in (b) the chain structures result in a small separation between dielectric cross-correlation and the hydrogen network from the $\alpha$ relaxation.

This results for glycerol rich concentrations in the coupling of $\alpha$ relaxation with $\tau_J$, $\tau_{\epsilon}$ and T$_g^{cal}$ and in propanol-rich concentrations in a small difference of $\alpha$ relaxation with $\tau_J$ and T$_g^{cal}$. Since $T_{\textbf{g}}$ is mainly focused on the $\alpha$ relaxation, the hydrogen-bonded superstructures are not visible but dominant in dielectric spectroscopy.\cite{huth2007comparing} These structures have been seen in propanol in scattering methods in the form of so-called pre-peaks \cite{Sillren:2013}. The same is not seen in Glycerol \cite{rucsax}. Additional X-ray measurements are planned to further understand the system's structure.

A good system to compare the investigated mixture to is glycerol/water \cite{jensen2018slow}, which consists of two H-bonded network forming substances with an $\Delta T_{\textbf{g}}$\,=\,190\,K-\,136\,K\,=\,54\,K instead of a network and chain-forming propanol/glycerol mixture with $\Delta T_{\textbf{g}}$\,=92\,K. In this system, we cannot observe an extreme spectral broadening and dynamic heterogeneity like in propanol/glycerol mixtures. Quite the opposite, the shear mechanical rheological spectra of glycerol lose their bimodality with increasing water concentration until the system starts to crystallize.

A different network-chain system to compare could be water/propanol ($T_{\textbf{g}}$ ratio 136\,K/100\,K) with a low $T_{\textbf{g}}$ difference of 36\,K. Data at high temperatures and high frequency does not show any suspicious broadening, quantified by Cole-Davidson stretching parameters between 1 and 0.9. A continuous decrease of relaxation strength from propanol to water could be attributed to a decreasing Kirkwood factor \cite{sato2000composition,sato2003dielectric}.


\section{Conclusions}

Glycerol and propanol mix well and were investigated using dielectric spectroscopy, mechanical spectroscopy, and calorimetry. All three methods support a large separation in dynamics while having additional relaxation in between. Features of hydrogen-bonded chain and network structures are still present in all BDS spectra at intermediate concentrations. The change in stiffness is evident from the mechanical measurement and can be interpreted as a percolation transition around $\chi_{gly}$ 0.5. An interpretation is that the hydrogen-bonded propanol chain structure becomes increasingly linked by glycerol, and that a network structure is constantly growing till percolation suddenly increases in stiffness. Propanol can move nearly freely up to a concentration of $\chi_{gly}=0.5$. The glycerol is accelerated by the propanol, and the overall dynamics are broadly distributed. This is a stark contrast to a water/glycerol mixture, where the dynamic is not separating and only a small broadening is observed \cite{jensen2018slow}. A rationalization for this behavior could be that water/glycerol is a network/network former mixture, whereas propanol/glycerol is a chain/network former mixture. Future investigations should reveal the possible influence of nano-structuring by investigating changes in pre-peaks in X-ray scattering.


\section{Supplementary Material}
Viscosity representations - time constant ratios between the permittivity and compliance - real parts of dielectric permittivity, temperature, and T$_g$ scaled - all complex shear and dielectric spectra.

\section{Acknowledgments}
I am grateful for the experimental support and discussions by Tina Hecksher and Jeppe Dyre, for X-ray Measurements with Dorthe Posselt. This work was supported by the VILLUM~Foundation’s \emph{Matter} grant (grant~no.~16515) and RUCSAXS – Roskilde University Interdisciplinary X-ray Scattering Hub – with grant NNF21OC0068491.

\bibliography{Literatur}

\widetext
\clearpage\begin{center}
\textbf{\large Supplementary Information:
Competing Supramolecular Structures: Dielectric and Rheological Spectroscopy on Glycerol/Propanol Mixtures}
\end{center}
\setcounter{equation}{0}
\setcounter{figure}{0}
\setcounter{table}{0}
\setcounter{page}{1}
\makeatletter

\subsection{Viscosity}

\begin{figure}[h!]
   \centering
   \includegraphics[width=7.5cm]{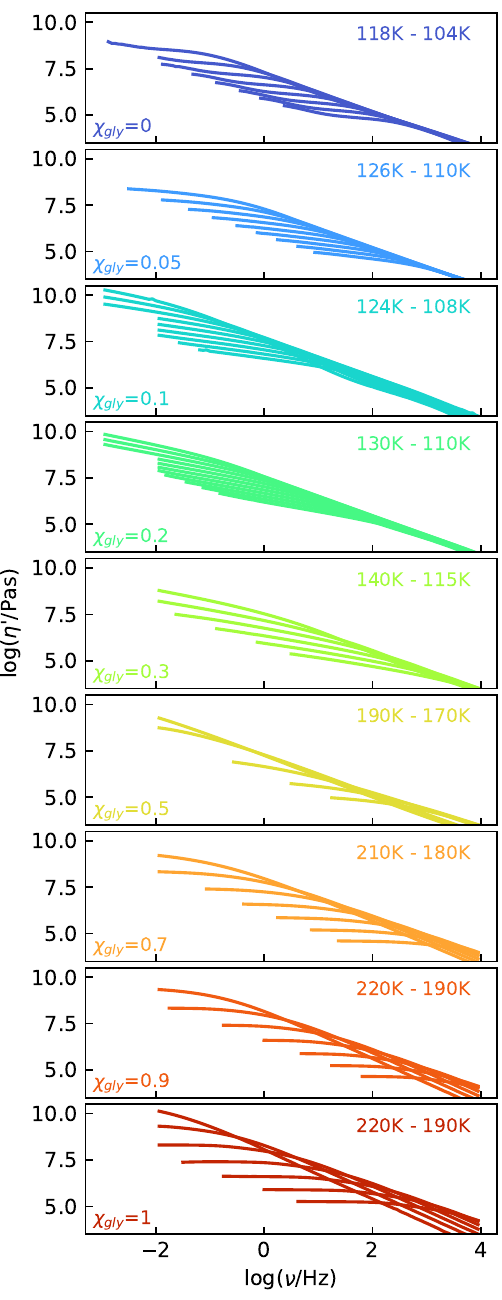}
   \caption{Viscosity representation - illustrates how close the measurements come to reaching a terminal flow.} 
\end{figure}

\begin{figure}[h!]
   \centering
   \includegraphics[width=8cm]{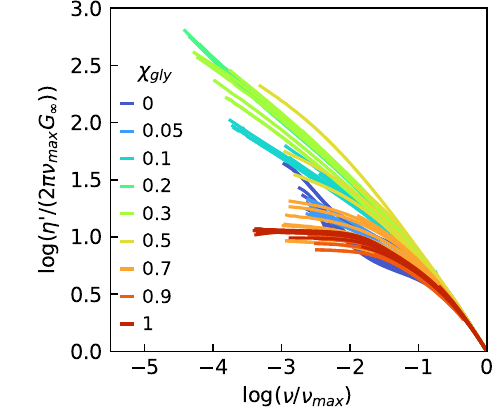}
   \caption{Scaled Viscosity representation - illustrates how close the measurements come to reaching a terminal flow - details \cite{bierwirth2018communication}.} 
\end{figure}

\subsection{Decoupling plot}

\begin{figure}[h!]
   \centering
   \includegraphics[width=8cm]{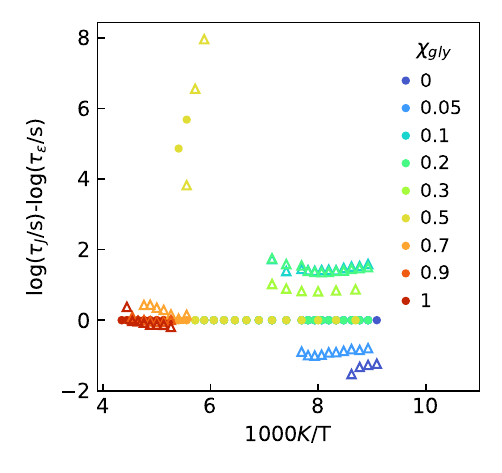}
   \caption{Decoupling plot - illustrates the ratio between the permittivity time constants and the main compliance time, as well as the secondary compliance time.} 
\end{figure}

\clearpage

\subsection{Real part dielectric permittivity}

\begin{figure}[h!]
   \centering
   \includegraphics[width=8cm]{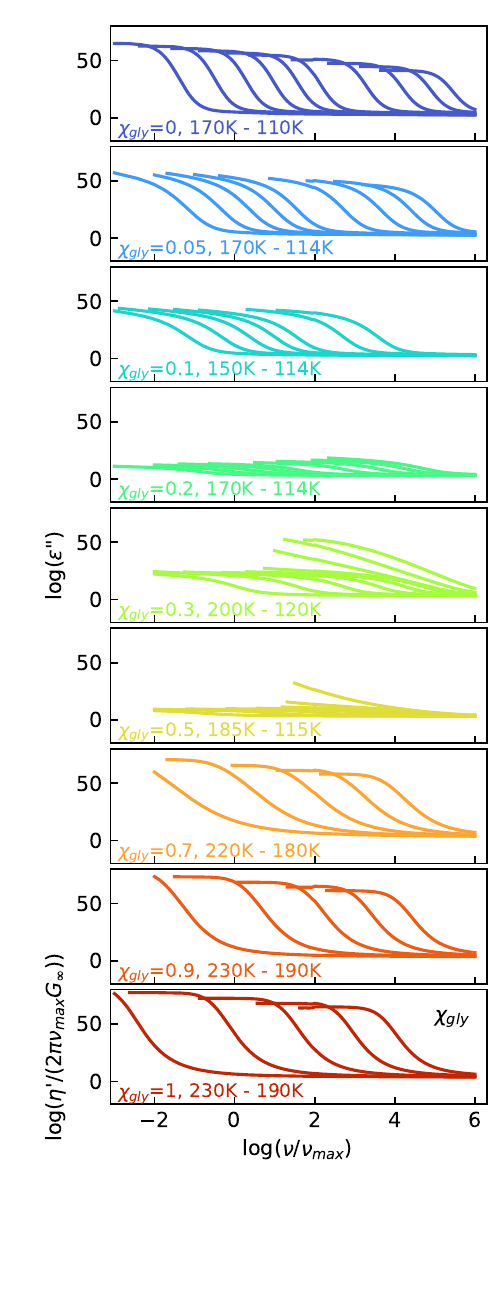}
   \caption{Concentration and temperature-dependent real part of the dielectric permittivity.} 
\end{figure}

\begin{figure}[h!]
   \centering
   \includegraphics[width=10cm]{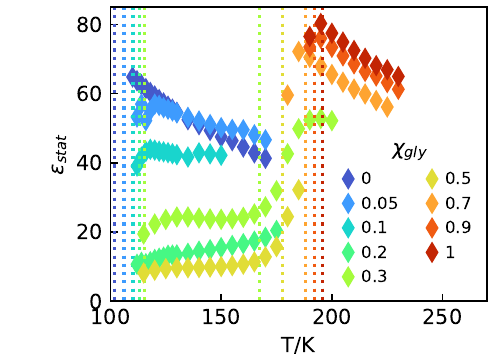}
   \caption{Concentration and temperature-dependent static permittivity. Dashed lines are the calorimetric T$_g$ values.} 
\end{figure}

\begin{figure}[h!]
   \centering
   \includegraphics[width=10cm]{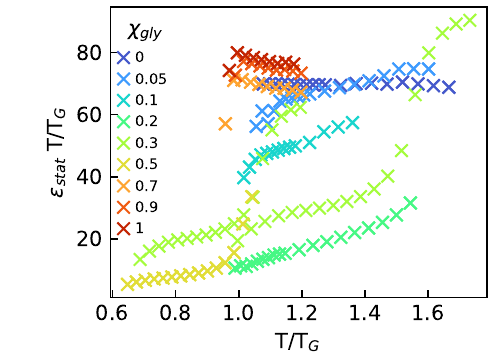}
   \caption{Concentration and temperature-dependent normalized static permittivity normalized to the calorimetric T$_g$. At a molar glycerol concentration of 0.3, the expected permittivity drop.} 
\end{figure}

\clearpage

\subsection{All shear spectra}

\begin{figure}[h!]
   \centering
   \includegraphics[width=9cm]{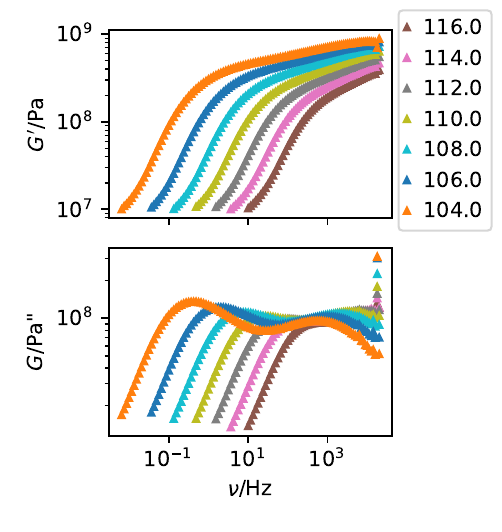}
   \caption{$\chi_{gly} = 0.0$} 
\end{figure}

\begin{figure}[h!]
   \centering
   \includegraphics[width=9cm]{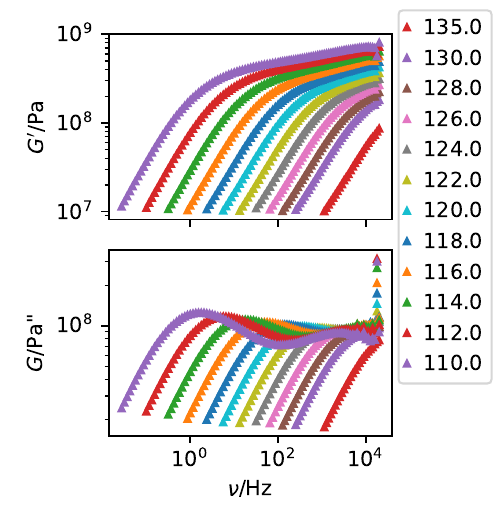}
   \caption{$\chi_{gly} = 0.05$} 
\end{figure}

\begin{figure}[h!]
   \centering
   \includegraphics[width=9cm]{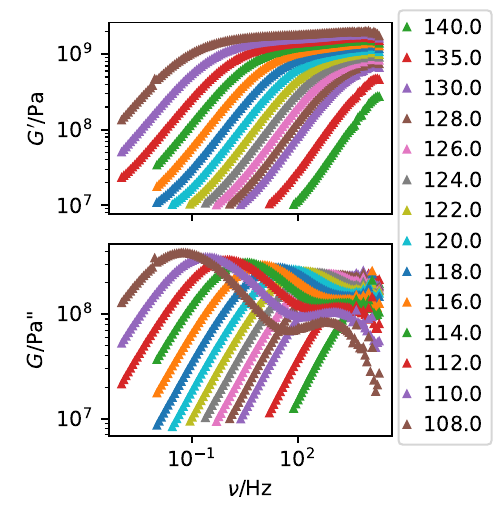}   
   \caption{$\chi_{gly} = 0.1$} 
\end{figure}

\begin{figure}[h!]
   \centering
   \includegraphics[width=9cm]{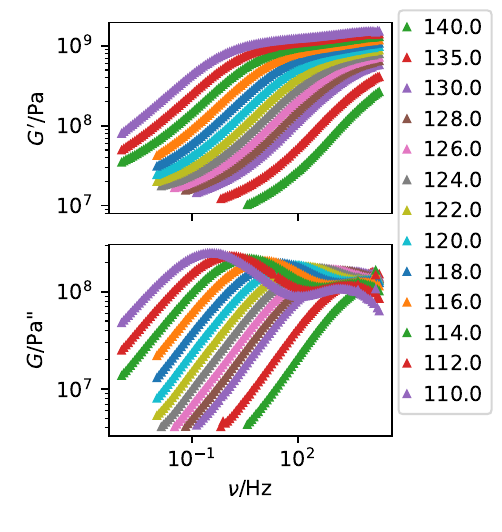}
   \caption{$\chi_{gly} = 0.2$} 
\end{figure}

\begin{figure}
   \centering
   \includegraphics[width=9cm]{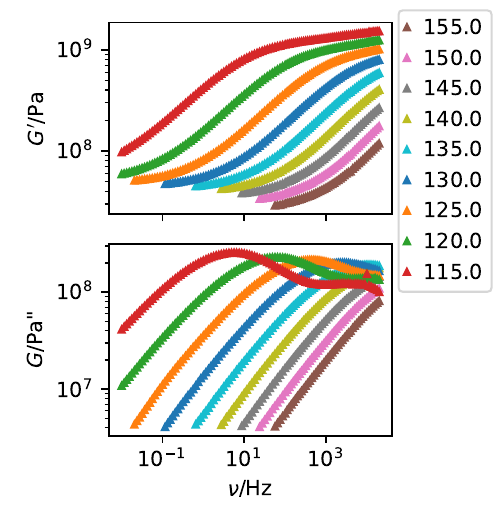}
   \caption{$\chi_{gly} = 0.3$} 
\end{figure}

\begin{figure}
   \centering
   \includegraphics[width=9cm]{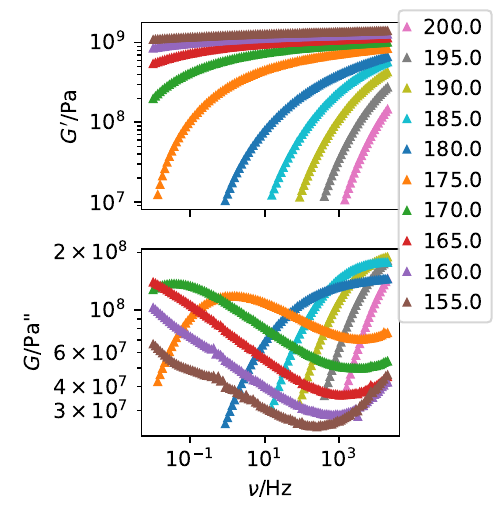}
   \caption{$\chi_{gly} = 0.5$} 
\end{figure}

\begin{figure}
   \centering
   \includegraphics[width=9cm]{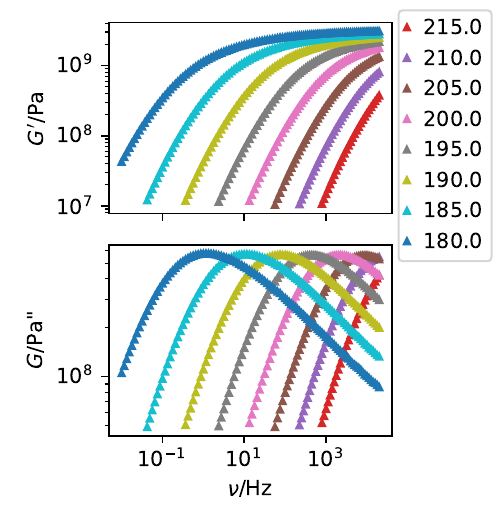}
   \caption{$\chi_{gly} = 0.7$} 
\end{figure}

\begin{figure}
   \centering
   \includegraphics[width=9cm]{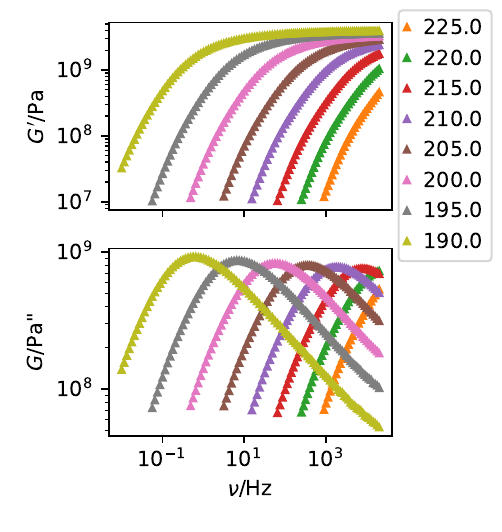}
   \caption{$\chi_{gly} = 0.9$} 
\end{figure}

\begin{figure}
   \centering
   \includegraphics[width=9cm]{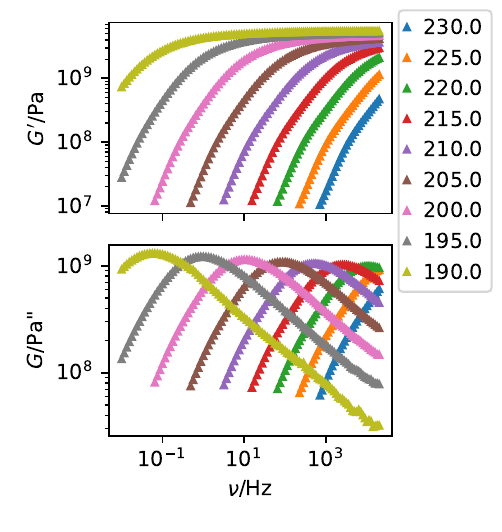}
   \caption{$\chi_{gly} = 1.0$} 
\end{figure}

\clearpage

\subsection{All dielectric spectra}

\begin{figure}[h!]
   \centering
   \includegraphics[width=9cm]{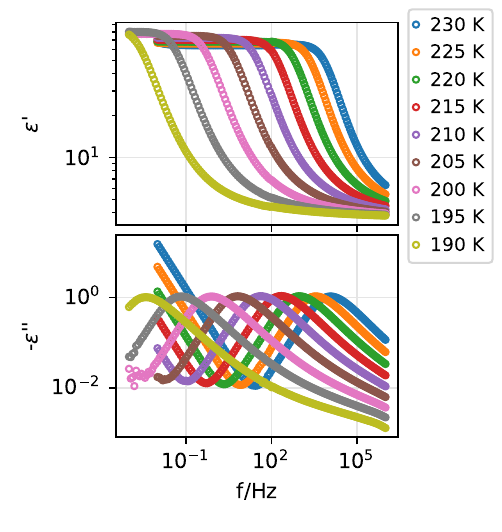}
   \caption{$\chi_{gly} = 1.0$} 
\end{figure}

\begin{figure}[h!]
   \centering
   \includegraphics[width=9cm]{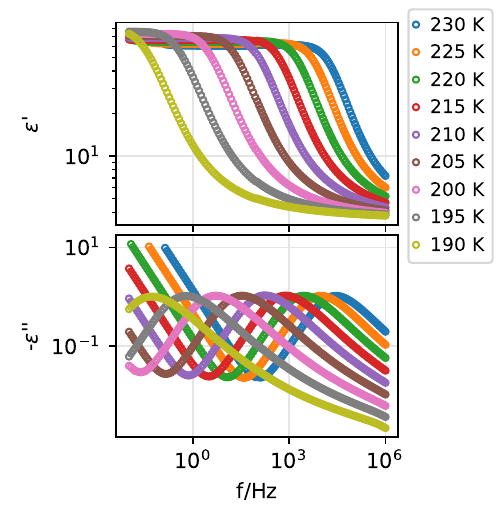}   
   \caption{$\chi_{gly} = 0.9$} 
\end{figure}

\begin{figure}[h!]
   \centering
   \includegraphics[width=9cm]{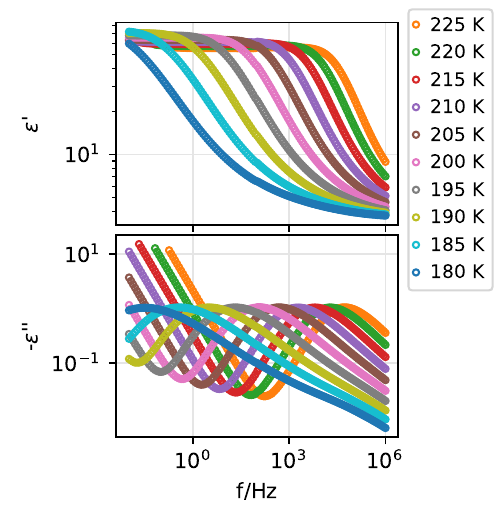}
   \caption{$\chi_{gly} = 0.7$} 
\end{figure}

\begin{figure}[h!]
   \centering
   \includegraphics[width=9cm]{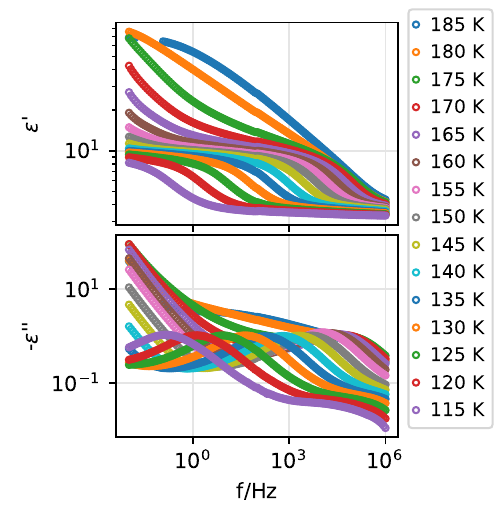}
   \caption{$\chi_{gly} = 0.5$} 
\end{figure}

\begin{figure}
   \centering
   \includegraphics[width=9cm]{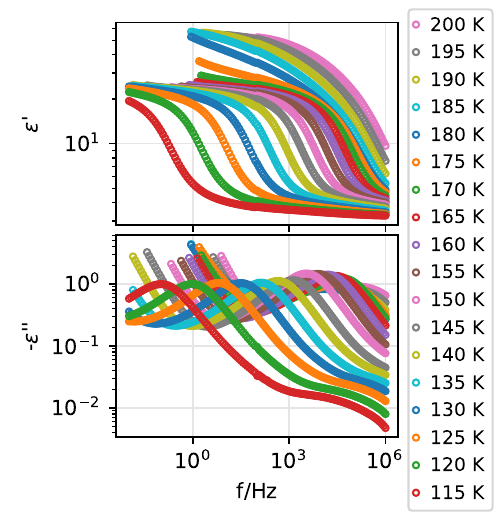}   
   \caption{$\chi_{gly} = 0.3$} 
\end{figure}

\begin{figure}
   \centering
   \includegraphics[width=9cm]{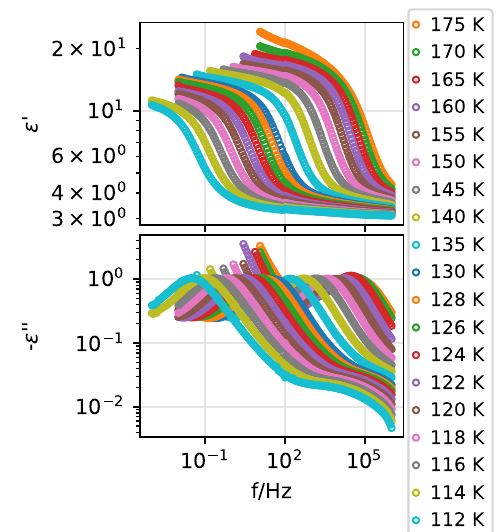}
   \caption{$\chi_{gly} = 0.2$} 
\end{figure}

\begin{figure}
   \centering
   \includegraphics[width=9cm]{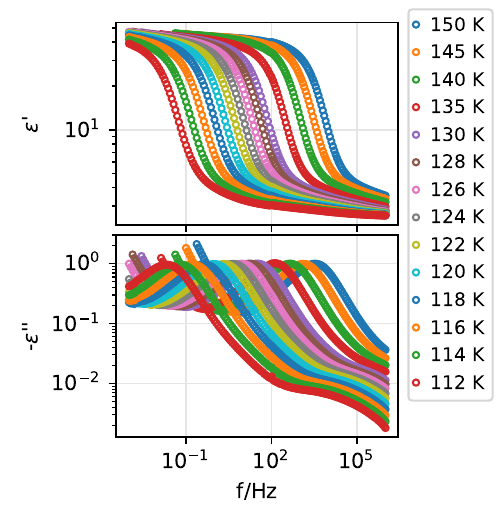}
   \caption{$\chi_{gly} = 0.1$} 
\end{figure}

\begin{figure}
   \centering
   \includegraphics[width=9cm]{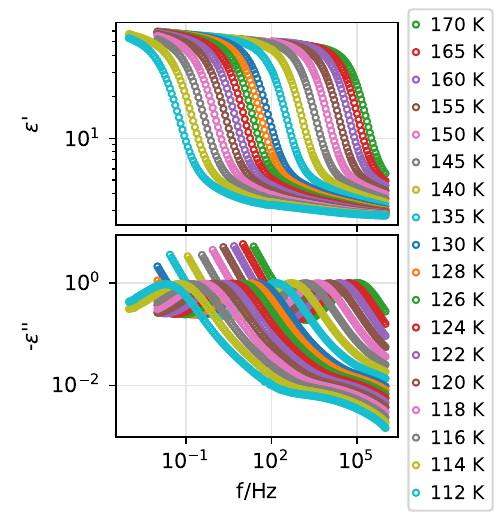}   
   \caption{$\chi_{gly} = 0.05$} 
\end{figure}

\begin{figure}
   \centering
   \includegraphics[width=9cm]{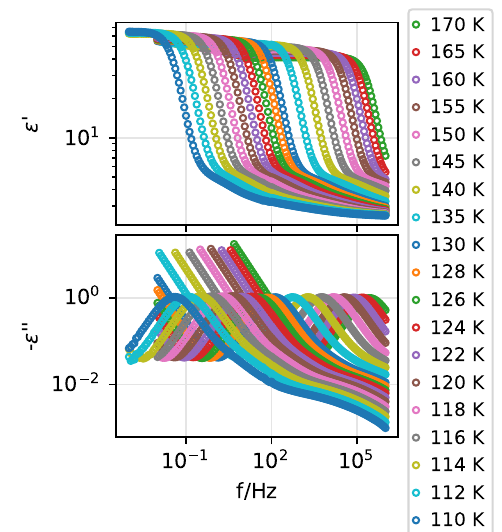}
   \caption{$\chi_{gly} = 0.0$} 
\end{figure}

\clearpage


\end{document}